\documentclass[12pt,thmsa,arabtex]{article}
\usepackage{amsfonts}
\usepackage{amsmath}
\usepackage{graphicx}

\input{tcilatex}
\begin{document}

\begin{center}
{\Large Sudden death and long-lived entanglement of two trapped ions }

Mahmoud Abdel-Aty$^{\ast,}$\footnote{E-mail:
abdelatyquantum@gmail.com\newline On leave from Mathematics
Department, Faculty of Science, Sohag University, 82524 Sohag,
Egypt} and H. Moya-Cessa$^{\ast \ast }$

$^{\ast }${\small Department of Mathematics, College of Science, University
of Bahrain, 32038, Kingdom of Bahrain}

$^{\ast \ast }${\small INAOE, Coordinacion de Optica, Apdo. Postal 51 y 216,
72000 Puebla, Pue., Mexico}

\bigskip
\end{center}

The dynamical properties of quantum entanglement in two effective two-level
trapped ions interacting with a laser field are studied in terms of the
negative eigenvalues of the partial transposition of the density operator.
In contrast to the usual belief that destroying the entanglement can be
observed due to the environment, it is found that the Stark shift can also
produce sudden death of entanglement and long-lived entanglement between the
qubits that are prepared initially in separable states or mixed states.

\bigskip

\section{Introduction}

As a promising resource, quantum entanglement plays a key role in quantum
information processing such as quantum teleportation \cite{rig05},
superdense coding \cite{ben92}, and quantum key distribution \cite{cur04}.
However, in the real world, quantum information processing will be
inevitably affected by the decoherence that destroys quantum superposition
and quantum entanglement. The extent to which decoherence affects quantum
entanglement is an interesting problem \cite{yu06}, and many researchers
study it extensively based on various models \cite{dor03,sub04,ma06,aso04}
with the point of view of environment induced decoherence.

The decay of entanglement cannot be restored by local operations and
classical communications, that is one of the main obstacles to achieve a
quantum computer \cite{ben96}. Therefore it becomes an important subject to
study the loss of entanglement \cite{yu06,yon06,pin06,yu02,yu03,yu04}. Quite
recently, by using vacuum noise two-qubit, entanglement terminated abruptly
in finite time has been performed \cite{yu06} and the entanglement dynamics
of a two two-level atoms model have been discussed \cite{yon06,li06}. They
called the non-smooth finite-time decay entanglement sudden death. Although
entanglement can be realized in different ways in experiments, how we can
preserve it is still a big challenge for current technology \cite{nie00}.
Because for open system, entanglement is fragile and decays exponentially,
it is often thought as similar as quantum decoherence. Most of the authors
who have treated this problem have dealt with the case in which the Stark
shift has been ignored \cite{yu06,yu06b}. However, in reality it cannot be
ignored. The main aim of the present paper is try to answer the following
question: what happens to two qubits entanglement when we consider the Stark
shift and different initial state setting in the absence or presence of the
decoherence?

We present an explicit connection between the initial state setting, Stark
shift and the dynamics of the entanglement. We give a condition for the
existence of either entanglement sudden death or long-lived entanglement. In
particular, a quantitative characterization of a general system of two
three-level trapped ions interacting with a laser field is presented. We
present various numerical examples in order to monitor the partial transpose
of the density operator and entanglement dynamics. In principle, by proper
adjustment of the initial state parameters, we can always find suitable
values of Stark shift which can be use to suppress the decay of entanglement.

\section{Model}

The physical system on which we focus is two effective two-level
harmonically trapped ions with their center-of-mass motion quantized. The
electronic levels $\left\vert a\right\rangle ,$ $\left\vert b\right\rangle $
and $\left\vert c\right\rangle $ are assumed to be metastable and coupled
via a laser field of the form \textrm{\cite{whe83}}
\begin{equation}
E(\hat{x},t)=E_{0}\exp [i(\hat{k}.\hat{x}-\omega t+\phi (t))],
\end{equation}%
where $E_{0}$ is the strength of the electric field, $\hat{k}$ is the wave
vector of the driving laser field, $\hat{x}$ is the position operator
associated with the center-of-mass motion and $\omega $ is the laser
frequency. We denote by $\phi (t)$ the fluctuations in the laser phase.
Therefore, we can express the center-of-mass position in terms of the
creation and annihilation operators of the one-dimensional trap namely $\hat{%
x}=\sqrt{\hbar /(2m\omega _{s})}(\hat{a}^{\dagger }+\hat{a})=\Delta x(\hat{a}%
^{\dagger }+\hat{a}).$ We denote by $\hat{a}$ and $\hat{a}^{\dagger }$ the
annihilation and creation operators and $\omega _{s}$ is the vibrational
frequency related to the center-of-mass harmonic motion along the direction $%
\hat{x}.$ In the absence of the rotating wave approximation, the trapped
ions Hamiltonian may be written as
\begin{equation}
\hat{H}=\hat{H}_{0}+\hat{H}_{1},  \label{1}
\end{equation}%
where $\hat{H}_{0}=\hbar \omega _{s}\hat{a}^{\dagger }\hat{a}+\hbar \omega
_{a}|a\rangle \langle a|+\hbar \omega _{b}|b\rangle \langle b|+\hbar \omega
_{c}|c\rangle \langle c|,$ and
\begin{equation}
\hat{H}_{1}=\hbar \sum\limits_{i=1}^{2}\left( \lambda _{1}^{(i)}e^{-i(k_{i}.%
\hat{x}-\omega t+\phi _{i})}S_{bc}^{(i)}+\lambda _{2}^{(i)}e^{-i(k_{i}.\hat{x%
}-\omega t+\phi _{i})}S_{ac}^{(i)}+c.c.\right) .  \label{2}
\end{equation}%
We denote by $\lambda _{1}^{(i)}=\left\langle b|\wp ^{(i)}|c\right\rangle
E_{01}$ $\ (\lambda _{2}^{(i)}=\left\langle a|\wp ^{(i)}|c\right\rangle
E_{02})$ the Rabi frequency characterizing the coupling strength (products
of dipole matrix elements and amplitudes of the incoming fields), where $\wp
^{(i)}$ is the dipole moment operator. As usual, to describe this system we
use the operators $S_{lm}^{(i)}=|l^{(i)}\rangle \langle m^{(i)}|,(l,m=a,b,c$
$\ $and $i=1,2).$ For the sake of simplicity (but without loss of
generality), we have assumed to deal with the case in which $\phi _{1}=0,$ $%
\phi _{2}=\phi $ and the level $\left\vert c\right\rangle $ is assumed to be
dipole-coupled to both the levels $\left\vert a\right\rangle $ and $%
\left\vert b\right\rangle $ via a far detuned laser field. While this is
straightforward, it is often the case that it is simpler to work in the
interaction picture in which the Hamiltonian (\ref{1}) evolves in time
according to the interaction with the vacuum field. If we express the
center-of-mass position in terms of the creation and annihilation operators,
the interaction part of equation (\ref{1}) becomes
\begin{eqnarray}
\hat{H}_{int} &=&\hbar \Delta \sum\limits_{i=1}^{2}\left(
S_{bb}^{(i)}+S_{aa}^{(i)}\right)   \notag \\
&&+\hbar \left( \lambda _{1}^{(1)}e^{-i\eta (\hat{a}^{\dagger }+\hat{a}%
)}S_{bc}^{(1)}+\lambda _{2}^{(1)}e^{-i\eta (\hat{a}^{\dagger }+\hat{a}%
)}S_{ac}^{(1)}+c.c.\right)   \notag \\
&&+\hbar \left( \lambda _{1}^{(2)}e^{-i\eta (\hat{a}^{\dagger }+\hat{a}%
)}S_{bc}^{(2)}e^{-i\phi }+\lambda _{2}^{(2)}e^{-i\eta (\hat{a}^{\dagger }+%
\hat{a})}S_{ac}^{(2)}e^{-i\phi }+c.c.\right) ,  \label{ham2}
\end{eqnarray}%
where $\eta =k\sqrt{\frac{\hbar }{2M\omega _{s}}},$ is the Lamb-Dicke
parameter and $\Delta $ is the detuning. In equation (\ref{ham2}) the
time-dependent factor is eliminated in the interaction picture, since $%
\omega _{c}-(\omega _{b}+\Delta )=\omega _{s}$ and $\omega _{c}-(\omega
_{a}+\Delta )=\omega _{s}$ (degenerate levels)$.$ Making use of the special
form of Baker-Hausdorff theorem \cite{blo92} the operator $\exp [i\eta
(a^{\dagger }+a)]$ may be written as a product of operators i.e. $\exp
(i\eta (a^{\dagger }+a))=\exp \left( \frac{\eta ^{2}}{2}[a^{\dagger
},a]\right) \exp \left( i\eta a^{\dagger }\right) \exp \left( i\eta a\right)
$ and assume the Lamb-Dicke regime with small $\eta $. In order to obtain
this we detune the laser frequency $\omega $ to the first vibrational red
sideband. Also, we apply the rotating wave approximation discarding the
rapidly oscillating terms and selecting the terms that oscillate with
minimum frequency \cite{tit96}. In these limits we can expand the
interaction Hamiltonian to lowest order in $\eta .$ The resulting
Hamiltonian may be written as
\begin{eqnarray}
\hat{H}_{int} &=&\hbar \Delta
\sum\limits_{i=1}^{2}(S_{bb}^{(i)}+S_{aa}^{(i)})+\hbar (\zeta _{1}^{(1)}\hat{%
a}^{\dagger }S_{bc}^{(1)}+\zeta _{2}^{(1)}\hat{a}^{\dagger
}S_{ac}^{(1)}+c.c.)  \notag \\
&&+\hbar (\zeta _{1}^{(2)}\hat{a}^{\dagger }S_{bc}^{(2)}e^{-i\phi }+\zeta
_{2}^{(2)}\hat{a}^{\dagger }S_{ac}^{(2)}e^{-i\phi }+c.c.),
\end{eqnarray}%
with a new coupling parameter $\zeta _{i}^{(j)}$ that includes the Dicke
parameter in its definition. The analysis of such a Hamiltonian model can be
carried out, providing eliminating of the non-resonantly coupled atomic
level $\left\vert c\right\rangle $ adiabatically, due to the large detuning,
the transitions for instance from the level $\left\vert a\right\rangle $ to
the level $\left\vert c\right\rangle $ are very fast and immediately
followed by decays on the atomic level $\left\vert b\right\rangle $.
Therefore, considering only coarse grained observables, meaning that the
system is observed at a rough enough time scale, effectively eliminates the
far detuned level, namely, at such a time scale, the only observables and
hence meaningful dynamical behaviors, involve levels $\left\vert
a\right\rangle $ and $\left\vert b\right\rangle $ as a result of time
averaging second order processes having $\left\vert c\right\rangle $ as an
intermediate virtual level. This procedure then suppresses the fine
dynamics, that is it sacrifices any information concerning the fast dynamics
the third level is involved in. So that the effective Hamiltonian of the
system including the ac-Stark shift, in the dipole and rotating wave
approximation, can be written as \cite{kud93,oba05} $(\hbar =1)$
\begin{eqnarray}
\hat{H}_{eff.} &=&\hat{a}^{\dagger }\hat{a}(\beta _{1}S_{bb}^{(1)}+\beta
_{2}S_{aa}^{(1)})+\hat{a}^{\dagger }\hat{a}(\beta _{1}S_{bb}^{(2)}+\beta
_{2}S_{aa}^{(2)})  \notag \\
&&+\zeta _{1}\left( S_{ab}^{(1)}\hat{a}^{2}+S_{ba}^{(1)}\hat{a}^{\dag
2}\right) +\zeta _{2}\left( e^{i\phi }S_{ab}^{(2)}\hat{a}^{2}+e^{-i\phi
}S_{ba}^{(2)}\hat{a}^{\dag 2}\right) .
\end{eqnarray}%
We denote by $\beta _{1}$ and $\beta _{2}$ the intensity-dependent Stark
shifts $\beta _{1}=\zeta _{_{1}}^{2}/\Delta ,$ and $\beta _{2}=\zeta
_{_{2}}^{2}/\Delta ,$ that are due to the virtual transitions to the
intermediate relay level and $\zeta _{i}=\frac{\zeta _{1}^{(i)}\zeta
_{2}^{(i)}}{\Delta },$ ($\Delta \neq 0$)$.$ This means that the two
three-level trapped ions (one-photon transitions) can be described by by an
effective two two-level system (in this case two-photon process).  \textrm{A
scheme utilizing position-dependent ac Stark shifts for doing quantum logic
with trapped ions has been presented \cite{sta02}. It has been shown that
specific ac Stark shifts can be assigned to the individual ions using a
proper choice of direction, position, and size, as well as power and
frequency of a far-off-resonant laser beam. }

The time evolution of the system density operator $\hat{\rho}(t)$ can be
written as \cite{bre02,lid03,gar00}
\begin{equation}
\frac{d}{dt}\hat{\rho}(t)=-\frac{i}{\hbar }[\hat{H},\hat{\rho}]-\frac{\gamma
}{2\hbar ^{2}}[\hat{H},[\hat{H},\hat{\rho}]],  \label{mas}
\end{equation}%
where $\gamma $ is the phase decoherence rate. Equation (\ref{mas}) reduces
to the ordinary von Neumann equation for the density operator in the limit $%
\gamma \rightarrow 0.$ The equation with the similar form has been proposed
to describe the intrinsic decoherence \cite{mil91}. Under Markov
approximations the \ solution of the master equation can be expressed in
terms of Kraus operators \cite{yu06} as follows%
\begin{eqnarray}
\hat{\rho}(t) &=&\sum_{m=0}^{\infty }\frac{\left( \gamma t\right) ^{m}}{m!}%
\hat{H}^{m}\exp \left( -i\hat{H}t\right) \exp \left( -\frac{\gamma t}{2}\hat{%
H}^{2}\right) \hat{\rho}(0)\exp \left( -\frac{\gamma t}{2}\hat{H}^{2}\right)
\exp \left( i\hat{H}t\right) \hat{H}^{m}  \notag \\
&=&\sum_{m=0}^{\infty }\frac{\left( \gamma t\right) ^{m}}{m!}\hat{M}^{m}(t)%
\hat{\rho}(0)\hat{M}^{\dagger m}(t),\qquad \qquad \qquad  \label{dens}
\end{eqnarray}%
where $\hat{\rho}(0)$ is the density operator of the initial state of the
system and $\hat{M}^{m}$ are the Kraus operators which completely describe
the reduced dynamics of the qubits system,
\begin{equation}
\hat{M}^{m}=\hat{H}^{m}\exp (-i\hat{H}t)\exp \left( -\frac{\gamma t}{2}\hat{H%
}^{2}\right) .
\end{equation}%
Equation (\ref{dens}) can also be written as
\begin{equation*}
\hat{\rho}(t)=\exp \left( -i\hat{H}t\right) \exp \left( -\frac{\gamma t}{2}%
\hat{H}^{2}\right) \{e^{\hat{S}_{M}t}\hat{\rho}(0)\}\exp \left( -\frac{%
\gamma t}{2}\hat{H}^{2}\right) \exp \left( i\hat{H}t\right)
\end{equation*}%
where we have defined the superoperator $\hat{S}_{M}\hat{\rho}(0)=\hat{H}%
\hat{\rho}(0)\hat{H}$. \textrm{We will choose the following mixed state of
the ions}
\begin{equation}
\rho ^{a}(0)=\cos ^{2}\theta \left\vert a,b\right\rangle \left\langle
a,b\right\vert +\sin ^{2}\theta \left\vert b,a\right\rangle \left\langle
b,a\right\vert \in \mathfrak{S}_{A}.  \label{6}
\end{equation}%
while the initial state of the vibrational mode is in a vacuum state $\rho
^{f}(0)=|0\rangle \langle 0|\in \mathfrak{S}_{F}.$ Then the initial state of
the system can be written as $\hat{\rho}(0)=\rho ^{a}(0)\otimes \rho
^{f}(0). $

\section{Entanglement}

In this paper we take the measure of negative eigenvalues for the partial
transposition of the density operator. It was proved that the negativity is
an entanglement monotone \cite{vid02}, hence, the negativity is a good
entanglement measure. According to the Peres and Horodecki's condition for
separability \cite{per96}, a two-qubit state for the given set of parameter
values is entangled if and only if its partial transpose is negative. The
measure of entanglement can be defined in terms of the negative eigenvalues
of the partial transposition in the following form \cite{lee00}
\begin{equation}
I_{\rho }\left( t\right) =2\max \left( 0,-\lambda _{neg}^{(i)}\right)
\end{equation}%
where $\lambda _{neg}^{(i)}$ is the sum of the negative eigenvalues of the
partial transposition of the time-dependent reduced atomic density matrix $%
\rho ^{a},$ which can be obtained by tracing out the vibrational mode
variables
\begin{equation}
\rho ^{a}=Tr_{f}\left( \hat{\rho}(t)\right) ,
\end{equation}%
In the two qubit system ($C^{2}\otimes C^{2})$ it can be shown that the
partial transpose of the density matrix can have at most one negative
eigenvalue \cite{per96}. The entanglement measure then ensures the scale
between $0$ and $1$ and monotonously increases as entanglement grows. An
important situation is that, when $I_{\rho }\left( t\right) =0$ the two
qubits are separable and $I_{\rho }\left( t\right) =1$ indicates maximum
entanglement between the two qubits. In our calculations, we have used the
two qubit basis $|aa\rangle ,|ab\rangle ,|ba\rangle $ and $|bb\rangle $ to
obtain the evolution of the density matrix of the system.

\begin{figure}[tbph]
\begin{center}
\includegraphics[width=10cm,height=6cm]{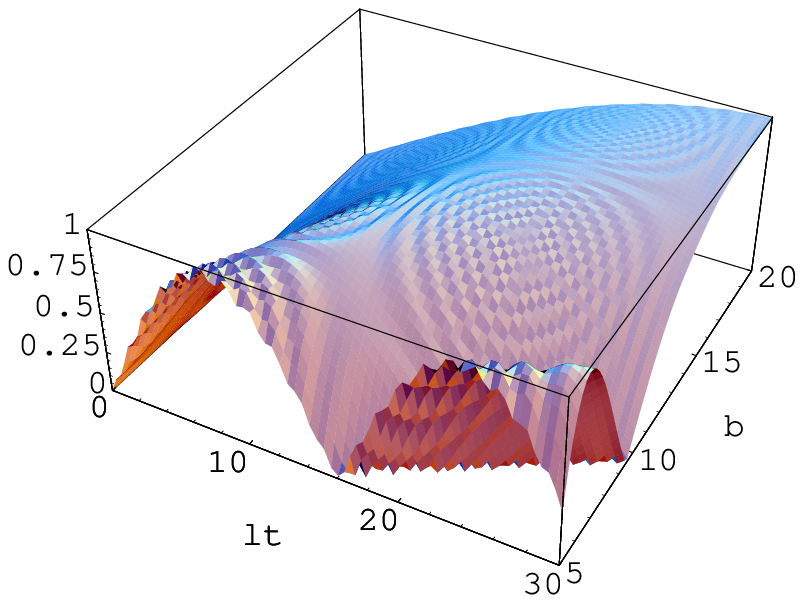} %
\includegraphics[width=5cm,height=5cm]{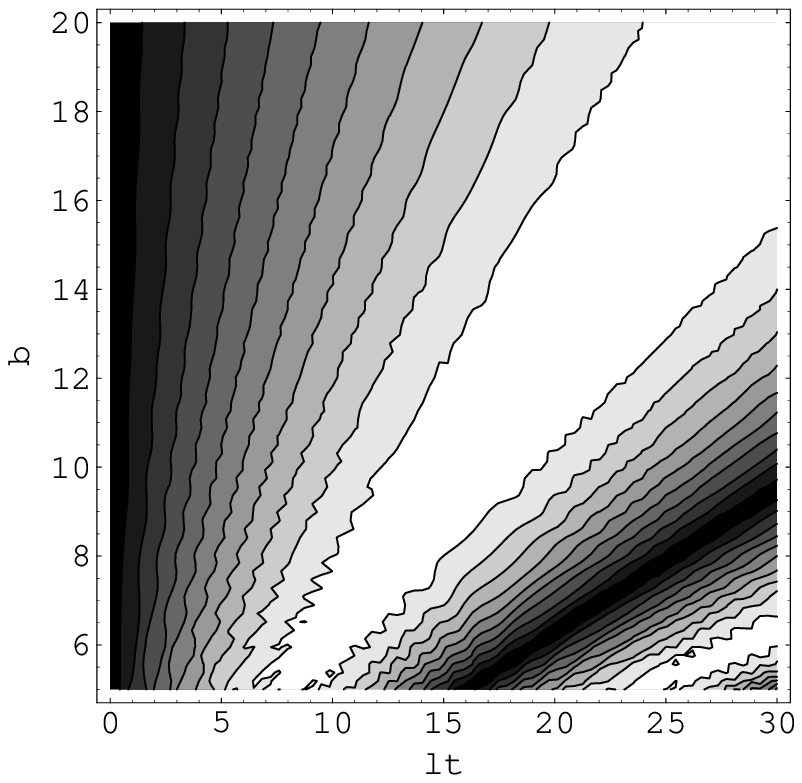}
\end{center}
\caption{The evolution of the quantum entanglement $I_{\protect\rho }\left(
t\right) $ as a function of the scaled time $\protect\lambda t,$ ($\protect%
\lambda =\protect\zeta _{1}=\protect\zeta _{2})$ and Stark shift parameter $%
\protect\beta .$ The parameters are $\protect\theta =0$ and $\protect\gamma %
=0$.}
\end{figure}

\textrm{An interesting question is whether or not the entanglement is
affected by the different parameters of the present system with the initial
state in which one of the qubits is prepared in its excited state and the
other in the ground state.} In particular, we focus on the effect of the
mixed state parameter $\theta $, the Stark shift parameter $\beta ,$ $%
(\equiv \beta _{1})$ and the decoherence. As expected from the results
presented in \cite{zhe00}, the analytical solution of equation (\ref{dens}),
when we set $\gamma =0,$ does not depend on the Stark shift parameter $%
\beta_{2}$, this result can be understood as coming from the setting of the
initial state of the vibrational mode, which was assumed to be in the vacuum
state.

A numeric evaluation of the entanglement measure leads to the plot in figure
1. We consider the initial state of the two ions $\theta =0.$ In this case,
we see that the entanglement is equal to zero in a periodic way for a small
values of the Stark shift parameter, this period is increased with
decreasing the parameter $\beta $. It is remarkable to see that with the
value of the Stark shift parameter, $\beta =20$ the entanglement is only
zero for the initial period of the interaction time, while long lived
entanglement is observed as the time goes on. In this case we can say that,
when the system is allowed to evolve without applying a phase shift ($\phi
=0 $), the entanglement is a periodic function of time for small values of
the Stark shift while long-survival entanglement can be obtained for larger
values of Stark shift (see figure 1). This is particularly because of the
nonlinear nature of the coupling in this case (two-photon process) \cite%
{zho02}.

\begin{figure}[tbph]
\begin{center}
\includegraphics[width=14cm,height=4.5cm]{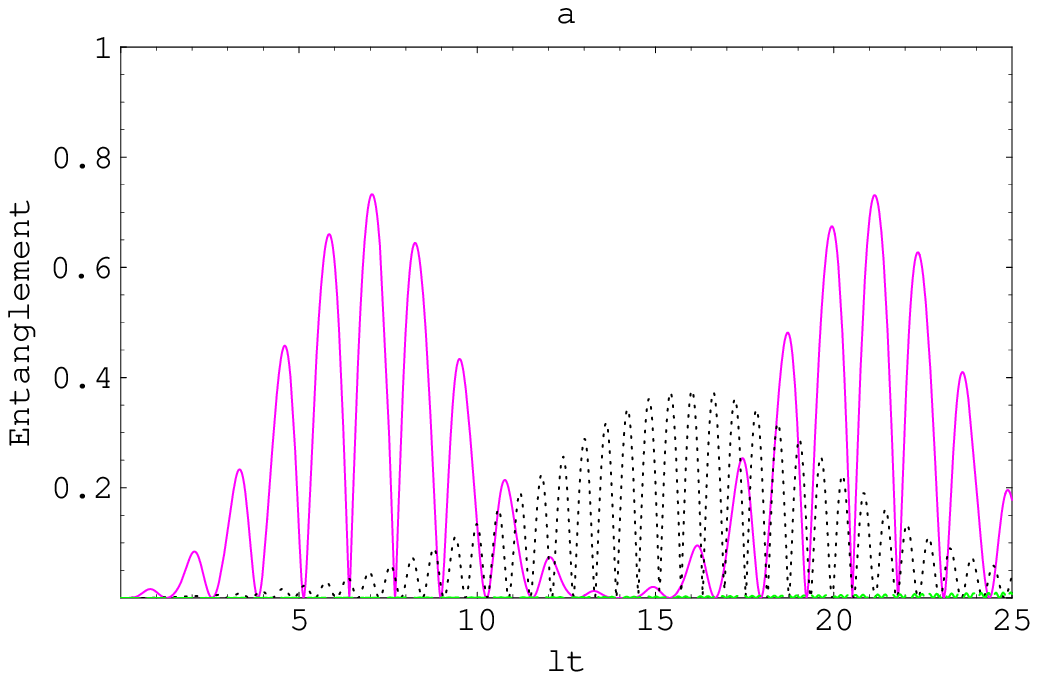} %
\includegraphics[width=14cm,height=4.5cm]{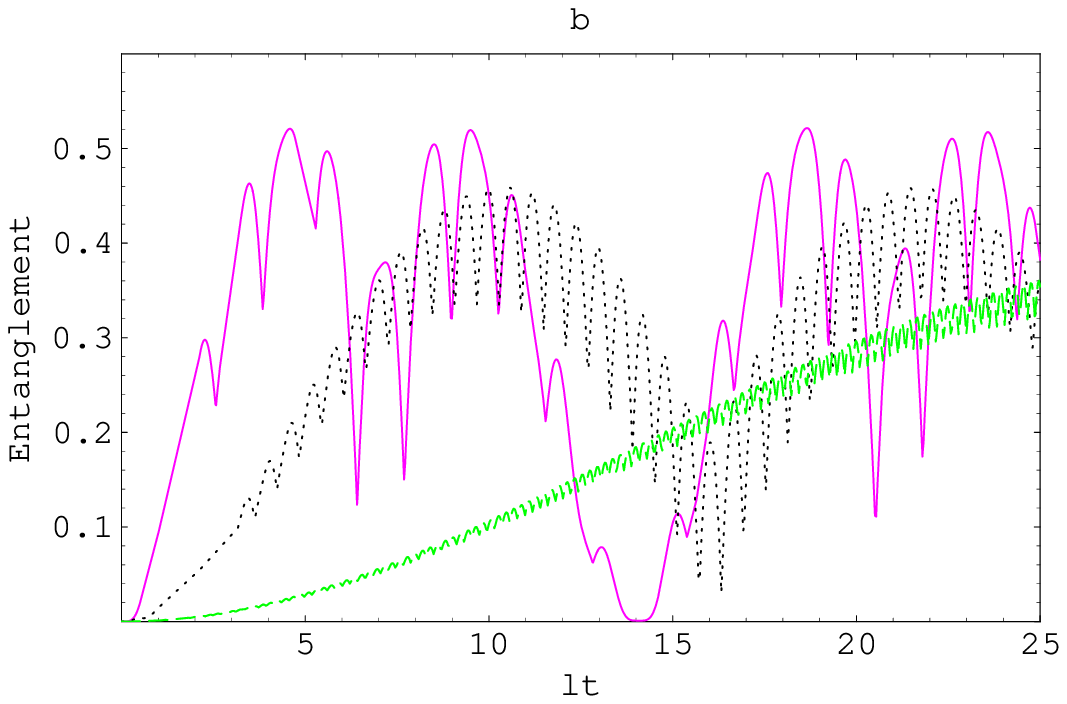}
\end{center}
\caption{The evolution of the quantum entanglement $I_{\protect\rho }\left(
t\right) $ as a function of the scaled time $\protect\lambda t$ and
different values of Stark shift parameter $\protect\beta$, where, $\protect%
\beta=2$ (solid curve), $\protect\beta=5$ (dotted curve) and $\protect\beta%
=15$ (dashed curve). The parameters are $\protect\gamma =0$ and (a) $\protect%
\theta =\frac{\protect\pi }{2}$ and (b) $\protect\theta =\frac{\protect\pi}{4%
}$.}
\end{figure}

We see from the figure 2a that large values of the Stark shift parameter
leads to zero entanglement. The situation here is quite different from that
observed in figure 1$,$ where the entanglement exist only for small values
of Stark shift while vanishes for all periods of the interaction time when $%
\beta >12.$ In figure 2b, we pause to touch on certain entanglement features
when a mixed state of the qubits is considered as $\rho
^{a}(0)=0.5(\left\vert a,b\right\rangle \left\langle a,b\right\vert
+\left\vert b,a\right\rangle \left\langle b,a\right\vert )$ i.e $\theta =\pi
/4.$ It is interesting to see here that the first maximum value of the
entanglement is observed at earlier time than the previous case. For large
values of the Stark shift parameter we see that the entanglement has zero
value only for a short period of the interaction time and then starts to
increase. This zero entanglement period is increased when the Stark shift in
increased further. These properties show that the role played by the Stark
shift on the entanglement is essential. Interestingly, when $\beta $ is
taken to be non zero, the values of the maximum entanglement are decreased,
indicating that the mixed state setting leads to a decreasing of the
qubit-qubit entanglement. Generally speaking, because of the influence of
mixed state parameter on entanglement, the amplitude of local maxima and
minima decrease with increasing the deviation of $\beta $ from the unity.
However, as $\beta $ takes values close to the unity we return to the same
behavior in the initial pure state setting i.e $\rho =\left\vert
a,b\right\rangle \otimes \left\langle a,b\right\vert .$ However a slight
change in $\beta $ therefore, dramatically alters the entanglement. This is
remarkable as the entanglement is strongly dependent on the initial state,
which can be entangled or unentangled.

We devote the discussion in figure 3 to consider the decoherence parameters
effect on the entanglement in the presence of Stark shift. We would like to
remark that decoherence due to normal decay is often said to be the most
efficient effect in physics. Which means that, the entanglement increases
rapidly, then approaches to a minimum value in a periodic manner. In this
case, the entanglement introduced by the coherent interaction oscillates
without dissipation, as showed in Fig. 3. Once, the environment has been
switched on, i.e., $\gamma \neq 0$, it is very clear that the decoherence
plays a usual role in destroying the entanglement. Also, from numerical
results we note that with the increase of the parameter $\gamma $, a rapid
decrease of the entanglement (entanglement sudden death) is shown (in
agreement with \cite{yu06}).

\begin{figure}[tbph]
\begin{center}
\includegraphics[width=14cm,height=4.5cm]{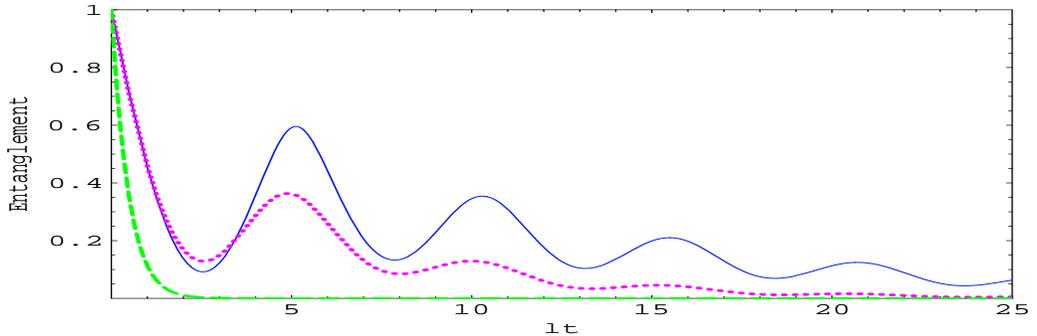}
\end{center}
\caption{The evolution of the quantum entanglement as a function of the
scaled time $\protect\lambda t$ and different values of the decoherence
parameter $\protect\gamma $, where, $\protect\gamma =0.01$ (solid curve), $%
\protect\gamma =0.1$ (dotted curve) and $\protect\gamma =0.7$ (dashed
curve). The other parameters are $\protect\theta =\protect\pi /2$ and $%
\protect\beta _{1}/\protect\beta _{2}=1$.}
\end{figure}

The remaining task is to identify and compare the results presented above
for the entanglement degree with another accepted entanglement measure such
as the concurrence \cite{vid02}. One, possibly not very surprising,
principal observation is that the numerical calculations corresponding to
the same parameters, which have been considered in figures 1-3, give nearly
the same behavior. This means that both the entanglement due to the
negativity and concurrence measures are qualitatively the same.

\section{Conclusion}

We have investigated the entanglement in the context of an ensembles of two
identical qubits (or ions) and negativity as computable measure of the
mixed-state entanglement has been used. We have treated the more general
case where the initial state of the two qubits can be mixed taking into
account the presence of Stark shift. Through analysis, we find that the
extent to which that the entanglement vanishes due to Stark shift relies not
only on the Stark shift value, but also on the initial state setting. When
the two ions start from a mixed state, the larger the Stark shift is, the
faster the entanglement vanishes. For pure quantum states, the complete
disentanglement occurs for a very short time in a periodic way only for
small values of the Stark shift. We found that, the entanglement decay due
to Stark shift for an initial mixed state is similar to the entanglement
decay due to the decoherence. Finally, we expect our work will be helpful
for preserving entanglement in practical experiments.

\section{Acknowledgements}

One of us (M A) would like to thank Prof. J. H. Eberly for his stimulating
communications.

\end{document}